\documentclass[reqno]{amsart}
\usepackage{color} 
\usepackage{times,amssymb,amsmath,amsfonts,amsthm,bbm,mathrsfs,color,enumitem,booktabs,cite,bm}
\usepackage[normalem]{ulem}
\usepackage{hyperref}
\usepackage[top=1.2in,bottom=1.2in,left=1.2in,right=1.2in]{geometry}
\allowdisplaybreaks
\newcommand\redout{\bgroup\markoverwith{\textcolor{red}{\rule[0.5ex]{2pt}{0.8pt}}}\ULon}

\newtheorem{theorem}{Theorem}
\newtheorem{lemma}[theorem]{Lemma}

\newtheorem{definition}{Definition}

\newcommand\nc\newcommand
\nc\ffa{{\boldsymbol a}}\nc\ffA{{\boldsymbol A}}\nc\cA{{\mathcal A}}
\nc\ffb{{\boldsymbol b}}\nc\ffB{{\boldsymbol B}}\nc\cB{{\mathscr B}}
\nc\ffc{{\boldsymbol c}}\nc\ffC{{\boldsymbol C}}\nc\cC{{\mathcal C}}
\nc\ffd{{\boldsymbol d}}\nc\ffD{{\boldsymbol D}}\nc\cD{{\mathscr D}}
\nc\ffe{{\boldsymbol e}}\nc\ffE{{\boldsymbol E}}\nc\cE{{\mathscr E}}
\nc\fff{{\boldsymbol f}}\nc\ffF{{\boldsymbol F}}\nc\cF{{\mathscr F}}
\nc\ffg{{\boldsymbol g}}\nc\ffG{{\boldsymbol G}}\nc\cG{{\mathscr G}}\nc\bG{{\mathbb G}}
\nc\ffh{{\boldsymbol h}}\nc\ffH{{\boldsymbol H}}\nc\cH{{\mathscr H}}
\nc\ffi{{\boldsymbol i}}\nc\ffI{{\boldsymbol I}}\nc\cI{{\mathcal I}}
\nc\ffj{{\boldsymbol j}}\nc\ffJ{{\boldsymbol J}}\nc\cJ{{\mathscr J}}
\nc\ffk{{\boldsymbol k}}\nc\ffK{{\boldsymbol K}}\nc\cK{{\mathscr K}}
\nc\ffl{{\boldsymbol l}}\nc\ffL{{\boldsymbol L}}\nc\cL{{\mathscr L}}
\nc\ffm{{\boldsymbol m}}\nc\ffM{{\boldsymbol M}}\nc{\cM}{{\mathscr M}}
\nc\ffn{{\boldsymbol n}}\nc\ffN{{\boldsymbol N}}\nc\cN{{\mathscr N}}
\nc\ffo{{\boldsymbol o}}\nc\ffO{{\boldsymbol O}}\nc\cO{{\mathscr O}}
\nc\ffp{{\boldsymbol p}}\nc\ffP{{\boldsymbol P}}\nc\cP{{\mathscr P}}
\nc\ffq{{\boldsymbol q}}\nc\ffQ{{\boldsymbol Q}}\nc\cQ{{\mathscr Q}}
\nc\ffr{{\boldsymbol r}}\nc\ffR{{\boldsymbol R}}\nc\cR{{\mathscr R}}
\nc\ffs{{\boldsymbol s}}\nc\ffS{{\boldsymbol S}}\nc\cS{{\mathscr S}}
\nc\fft{{\boldsymbol t}}\nc\ffT{{\boldsymbol T}}\nc\cT{{\mathscr T}}
\nc\ffu{{\boldsymbol u}}\nc\ffU{{\boldsymbol U}}\nc\cU{{\mathscr U}}
\nc\ffv{{\boldsymbol v}}\nc\ffV{{\boldsymbol V}}\nc\cV{{\mathscr V}}
\nc\ffw{{\boldsymbol w}}\nc\ffW{{\boldsymbol W}}\nc\cW{{\mathscr W}}
\nc\ffx{{\boldsymbol x}}\nc\ffX{{\boldsymbol X}}\nc\cX{{\mathscr X}}
\nc\ffy{{\boldsymbol y}}\nc\ffY{{\boldsymbol Y}}\nc\cY{{\mathscr Y}}
\nc\ffz{{\boldsymbol z}}\nc\ffZ{{\boldsymbol Z}}\nc\cZ{{\mathscr Z}}
\nc{\bb}{{\mathbbm{1}}}
\nc\reals{{\mathbb R}}
\nc{\ff}{{\mathbb F}}
\nc{\PP}{{\mathbb P}}
\nc{\integers}{{\mathbb Z}}

\newcommand{\floor}[1]{\left \lfloor #1 \right \rfloor}

\DeclareMathOperator{\rank}{rk}

\newcommand\remove[1]{}

\DeclareSymbolFont{bbold}{U}{bbold}{m}{n}
\DeclareSymbolFontAlphabet{\mathbbold}{bbold}

\begin{document}
\title{A construction of maximally recoverable codes}
\author{Alexander Barg}\address{Department of ECE and Institute for Systems Research, University of Maryland, College Park, MD 20742, USA and Inst. for Probl. Inform. Trans., Moscow, Russia}\email{abarg@umd.edu}
\author{Zitan Chen}\address{}\email{chenztan@gmail.com}
\author{Itzhak Tamo}\address{Department of EE-Systems, Tel Aviv University, Tel Aviv, Israel}\email{zactamo@gmail.com}

\begin{abstract} We construct a family of  linear maximally recoverable codes with locality $r$ and dimension $r+1.$ For codes of length $n$ with $r\approx n^\alpha, 0\le\alpha\le 1$ the code alphabet is of the order $n^{1+3\alpha},$ 
which improves upon the previously known constructions of maximally recoverable codes.
\end{abstract}

\maketitle
\section{Introduction}
Consider a linear code $\cC$ over a finite field $F=\ff_q$ of length $n$ and dimension $k$, and let $r$ be a number such 
that $r+1$ divides $n$. We will write $[n]=\{(i,j),j=1,\dots,r+1;\,i=1,\dots,\frac n{r+1}\}$, and for  
 $i=1,\dots,\frac n{r+1}$ we will call the subset of indices $R_i=\{(i,j), j=1,\dots,r+1\}$ a {\em repair group}. Call a set $T\subset[n]$ a transversal of the set of repair groups $\cR=(R_i)_i$ if $|T\cap R_i|=1$ for all $i$.
For a subset $X\subset [n]$ denote by $\cC|_X$ the puncturing of $\cC$ in the coordinates in $X$, i.e., a coordinate projection of $\cC$ on the complement of $X$ in $[n]$.

\begin{definition}\label{def:MRcode}  The code $\cC$ is called \emph{maximally recoverable} (MR) with locality $r$ if the following two properties hold:
\begin{itemize}
     \item[(i)] For any repair group $R_i$ the code $\cC|_{R_i^c}$ has distance at least $ 2$;
     \item[(ii)] For any transversal $T$ of $\cR$ the code $\cC|_T$ is maximum distance
separable.
\end{itemize}
We write the parameters of the code $\cC$ as $(n,k,r)=(\text{length,dimension,locality}).$
\end{definition}

This definition is a particular case of a more general notion of MR codes introduced in \cite{Gopalan2014}. Namely, one assumes that
(i), any repair group is capable of correcting any $a\ge1$ erasures, and (ii), upon puncturing any $a$ coordinates from each of
the repair groups, the punctured code is maximum distance separable that can correct $h$ erasures. Thus,  Definition \ref{def:MRcode}
corresponds to the case of $a=1$ and $h=\frac {nr}{r+1}-k.$ 

The main problems related to MR codes are: the minimum field size $q$ required to construct an MR code with a given set of parameters, and finding 
explicit constructions of MR codes, with sizeable literature devoted to them over the last decade.
In this note we prove the following  result.
\begin{theorem}\label{thm:main} There exists a family of $(n,k=r+1,r)$ MR codes over $\ff_q,$ where $r=\Theta(n^\alpha), 0\le \alpha\le 1$,  with 
   $
    q=
      \Theta(n^{1+3\alpha+o(1)}).
     $
\end{theorem}   
To establish it, we develop an idea behind one construction in \cite{Gopi2020} that gave codes with locality $r=3$ and large $k$ 
relying on Behrend's classic result on sets of integers with no 3-term arithmetic progressions \cite{Behrend1946}. To address the case of general $r$ we use Alon's extension of Behrend's construction \cite{Alon2001}.

 For the context we include a sample of the known results on the construction and parameters of MR codes, focusing on the regime of large $h$ relevant to us. Among the known families of MR codes we note constructions with $q=O((r+1) n^{2h-1})$ \cite{Gabrys2019}, $q=O((r+1)^{\frac{nr}{r+1}})$ \cite{Neri2020}, $q=O(\max(r+1,\frac n{r+1}))^h$ \cite{cai2020construction}, as well as a number of constructions in \cite{Guruswami2020} with comparable parameters. We refer the reader to the introduction of \cite{Gopi2020,Guruswami2020}, or \cite{cai2020construction}, Table 1, for a more detailed overview which also covers the entire range of possible parameters $a,r,$ and $h$. As remarked in \cite{Guruswami2020}, most of the known constructions require alphabets of size $q$ that depend exponentially on $h$. One exception where the minimum alphabet size is independent of $h$ is the construction of \cite{Martinez2019} which requires $q=O(\max(n/(r+1),r+1))^r.$ This is larger than the result in Theorem \ref{thm:main} above both for fixed and growing $r$.  
  In summary, the code family constructed in this paper improves upon the known results in terms of the required field size.

To address the question of lower bounds (impossibility), we observe that the known nontrivial results \cite{Gopi2020} assume that $h$ is a fixed constant. 
At the same time, in our setting $h$ is clearly increasing with $n$, and the only known constraints on $q$ are general 
bounds of the form $q=\Omega(n)$  (for instance, part (ii) of Definition \ref{def:MRcode} implies that $q\ge k+1$, see \cite[Theorem 19]{Gopalan2014}).
    
\section{The construction}
Let $\gamma$ be a primitive element of $F$ and let $N=q-1$ be the size of its multiplicative group. We will define a linear code over $F$ by specifying its generator matrix $G=(g_{\ell,(i,j)})$ of dimensions $(r+1)\times n$ where the length $n$ will be determined later. Let 
   $$
   \cA=\bigcup_{i=1}^{\frac{n}{r+1}} A_i\subset \integers_N
   $$
be a subset of size $n$ formed as a union of pairwise disjoint sets $A_i,$ and let $\{a_{ij},j=1,\dots,r+1\}$ be the elements of $A_i,
i=1,\dots,n/(r+1)$. Define
  \begin{equation}\label{eq:G}
   g_{\ell,(i,j)}=
\begin{cases}
\gamma^{\ell \cdot a_{ij}} & 1\leq \ell \leq r \\ 
\gamma^{(r+1)\cdot a_{ij}}+(-1)^{r+1} &  \ell= r+1. \\
\end{cases}
  \end{equation}
The main property of the set $\cA$ that supports the construction is the following: let $\{a_1,\dots,a_{r+1}\}\subset \cA,$ then
  \begin{equation}\label{eq:Z}
    \sum_{s=1}^{r+1}a_s=0 \;(\text{mod}\, N) \text{ if and only if $\exists$ } i\in\{1,\dots,\textstyle{\frac n{r+1}}\}
    \text{ s.t. }  A_i=\{a_1,\dots,a_{r+1}\}.
    \end{equation}

  \begin{theorem}
\label{thm:Code}
Let $G$ be the matrix defined above, where the set $\cA$ satisfies \eqref{eq:Z}. Then the rows of $G$ 
span an $(n,r+1,r)$-MR code over $F$. 
   \end{theorem}

Next we give a construction of the set $\cA$ with the required properties. Let $\lambda$ and $\delta$ satisfy
$0<\lambda<\frac{1}{r^3}, 0<\delta<\frac{\lambda}{r},$ and define $d=\floor{\delta N}, {l}=\floor{\lambda N}.$ Suppose that
$D\subset \{1,\ldots,d\}$ is a subset of integers such that for any $d_0,\dots,d_r\in D$ the equation over $\integers$
   \begin{equation}\label{eq:3ap}
   d_0+\dots+d_{r-1}=r d_r
   \end{equation}
is satisfied only if $d_0=d_1=\dots=d_r.$ Define $r+1$  subsets $D_i\subset \integers_N$ by letting
   \begin{equation}\label{eq:A}
   D_i=\begin{cases}
     i{l} +D & 0\leq i\leq r-1\\
     N-\binom{r}{2}{l}-r\cdot D & i=r,
       \end{cases}
  \end{equation}
where $b+D, b\cdot D$ mean adding or multiplying every element of $D$ by  $b.$ By the choice of $\lambda$ and $\delta$ one can verify that the subsets  $D_i$ are disjoint. 
Define the set $\cA=\cup_{i=0}^r D_i$ and note that $|\cA|=n:=|D|(r+1)$. 
Consider a partition of $\cA$ into disjoint transversals $A_b$ for any $b\in D$ where \begin{equation}\label{eq:Ai} 
A_b=\{a_{i,b}:i=0,\dots,r\}  \text{ and }
  a_{i,b}=\begin{cases}
    i {l}+b, &0\le i\le r-1\\
    N-\binom{r}{2}{l}-rb, &i=r.
   \end{cases}
  \end{equation}
   \begin{lemma}\label{thm:A} 
The partition $\cA=\cup_{b\in D} A_b$ satisfies property \eqref{eq:Z}.
   \end{lemma}

Large sets of integers that satisfy \eqref{eq:3ap} exist, namely, the following is true.
\begin{lemma}{\rm \cite[Lemma 3.1]{Alon2001}}\label{lemma:B}
For every $r\geq  2$ and every
positive integer $m$, there exists a subset $D \subset \{1, 2,\ldots, m\}$ of size at least
   $$
   |D| \geq  \frac{m}{{e}^{5\sqrt{\log m \log r}}}
   $$
that has property \eqref{eq:3ap}.
\end{lemma}
This claim is proved by an averaging argument over intersections of a subset of integers with spheres of varying radii, and this is the only non-explicit part of our construction. We include a short proof at the end of the next section to make the presentation self-contained.

Putting things together, we have constructed an $(n,r+1,r)$ MR code $\cC$ of length 
    $$
    n = |D|(r+1) \ge (r+1) d\, e^{-5\sqrt{\log d\log r}},
    $$
where we have used Lemma \ref{lemma:B} with $m=d.$ Let us estimate the dependence of the field size $q$ on the parameters of $\cC,$
letting $n,q\to \infty.$  
We have
   \begin{equation}\label{eq:n}
   \log n\ge \Omega\Big( \log  q-3\log r-5\sqrt{\log\frac q{r^4}\log r}\Big).
   \end{equation}
where we put $d=\Theta(\frac q{r^4})$ (this appears to be the best choice given our assumption on $\delta$). Suppose that $r= \Theta (n^\alpha),$ where $0\le \alpha\le 1$ and note that this 
includes the cases of constant $r$ and various rates of increase of $r$ up to $r=\Theta(n),$ i.e., a constant number of repair groups. Now from \eqref{eq:n} we obtain the estimate for $q$ stated in Theorem~\ref{thm:main}.

\section{Proofs}
\noindent{\em Proof of Theorem \ref{thm:Code}}: Let $S\subseteq [n]$ be an $(r+1)$-subset of 
indices and let $G_S$ be a square  submatrix of $G$ of order $r+1$ whose columns are indexed by the elements of $S$. 
We begin by showing that the rank of  $G_S$ is $r$ if  $S$ forms a repair group, otherwise, $G_S$ is of full rank. First note that the first $r$  rows of $G_S$ form   an $r\times (r+1)$ Vandermonde submatrix, hence the rank of $G_S$ is at least $r.$ The rank is exactly $r$ if and only if there exists a nonzero vector 
$f=(f_1,\ldots, f_{r},f_{r+1})$ such that $f\cdot G_S=0$. Note that $f_{r+1}\neq 0$ since otherwise, it would violate the fact that the first $r$ rows of $G_S$ are linearly independent. Therefore, assume wlog that $f_{r+1}=1$. Since the columns of $G$  are defined by  elements $\gamma^\beta, \beta\in \mathcal{A}$, the conditions $f\cdot G_S=0$ are alternatively written as $f(\gamma^{\beta_i})=0$ for $i=1,\cdots,r+1$ and some $\beta_i\in \mathcal{A}$, where
    $$
   f(x):=x^{r+1}+(-1)^{r+1}+\sum_{i=1}^{r}f_ix^i.
   $$
By assumption, the monic polynomial $f(x)$ has $r+1$ zeros $\gamma^{\beta_i}$, and thus
   $$
   f(x)=\prod_{i=1 }^{r+1}(x-\gamma^{\beta_i}).
   $$
By comparing the constant terms in the two expressions of  $f$  we have 
    $$
   (-1)^{r+1}=\prod_{i=1}^{r+1}(-\gamma^{\beta_i})=(-1)^{r+1}\gamma^{\sum_{i=1}^{r+1}\beta_i}
    $$ 
or  $\sum_{i=1}^{r+1} \beta_i=0 \;(\text{mod}\, N).$ Then, by recalling assumption \eqref{eq:Z}, either the subset $S$ forms a repair group, or otherwise $G_S$ is of full rank. Hence property (ii) in Definition \ref{def:MRcode} holds.  Next, assume that $S$ forms a repair group, and we need to show that  $\cC|_{S^c}$ has distance at least $2$. To prove this, note that any $(r+1)\times r$ submatrix of $G_S$ has rank $r$ since it contains an $r\times r$ Vandermonde submatrix. Since $\rank(G_S)=r,$ any column of $G_S$ is in the span of the remaining $r$ columns, and thus the code $\cC|_{S^c}$ corrects a single erasure. \qed

\vspace*{.15in}
\noindent{\em Proof of Lemma \ref{thm:A}}: Let $\cB:=\{b_0,b_1,\dots,b_r\}\subset \cA$.  We will show that 
\begin{equation}
\label{eq1}    
\sum_{i=0}^rb_i=0\;(\text{mod}\, N).
\end{equation}
is satisfied if and only if $\cB$ coincides with one of the transversals $A_b$ defined in \eqref{eq:Ai}.
One direction is easy: namely, the elements in every $A_b$ sum to $0$ modulo $N$. Indeed
   $$
   \sum_{j=0}^{r}a_{i,b}=\sum_{j=0}^{r-1}(i{l}+b)+N-\binom{r}{2}{l}-rb=0 \;(\text{mod}\, N).
   $$
Conversely, suppose \eqref{eq1} holds. We aim to prove that $\cB=A_b$ for some $b\in D.$
Let $t=|\cB \cap D_r|$ and let us first show that $t=1.$ 
Indeed, if $t=0$, then each  $b_i \notin D_r$ and therefore $b_i \le (r-1){l}+{d}$ over $\integers$, and (again over $\integers$)
   $$
   0< \sum_{i=0}^{r}b_i\leq r((r-1){l}+{d}) <r^2{l}<N,
   $$
where the last inequality follows by the choice of $\lambda.$
This contradicts \eqref{eq1}, so $t\ge 1.$  
A similar argument applies in the case of $t\geq 2,$
namely we will show that in such a case $(t-1)N < \sum_{i=0}^r b_i< tN$ over $\integers$, which again will contradict \eqref{eq1}. 
Indeed, we have
\begin{align*}
\sum_{i=0}^{r}b_i&>\sum_{b_i\in D_{r}}b_i\geq tN-t\binom{r}{2}{l}-tr{d}\geq (t-1)N+N\Big(1-t\lambda\Big(\binom{r}{2}+1\Big)\Big)\\
&=(t-1)N+N\Big(1-t\lambda\frac{r^2-r+2}{2}\Big)>(t-1)N,
\end{align*}
where the last step follows since $\lambda< r^{-3},t\leq r+1$ and $r\geq 2$.
For the upper bound write
\begin{align}
\sum_{i=0}^{r}b_i&= \sum_{b_i\in D_{r}}b_i +\sum_{b_i\notin D_{r}}b_i \nonumber\\
& \leq tN-t\binom{r}{2}{l}+(r+1-t)((r-1){l}+{d})\nonumber\\
& \leq tN-2\binom{r}{2}{l}+(r-1)((r-1){l}+{d})\nonumber\\
&= tN -(r-1)({l}-{d})<tN,\nonumber
\end{align}
again contradicting \eqref{eq1}. We conclude that $t=1$ and suppose that $D_{r}\cap \cB=\{b_r\}$, where
   \begin{equation}
\label{eq2}
b_{r}=N-\binom{r}{2}{l}-r{z},
    \end{equation} 
for some ${z}\in D.$ 

Our next goal is to show that all the other elements in $\cB$ are of the form $b_i=ie+{z},i=0,\dots,r-1,$ and here property \eqref{eq:3ap}
comes in handy. We begin with arguing that 
$t_i:=|\cB\cap D_i|=1$ for all $i=0,\ldots,r-1$. Note that over $\mathbb{Z}$ 
$$\sum_{i=0}^{r-1}b_i\leq r((r-1){l}+{d})<r^2{l}<N,$$
hence from \eqref{eq1} and \eqref{eq2} 
   $$
   \sum_{i=0}^{r-1}b_i=\binom{r}{2}{l}+r{z},
   $$ 
again over $\mathbb{Z}.$  Clearly, $\sum_{i=0}^{r-1}t_i=r,$ and we will show that
   \begin{equation}\label{eq:iti}
    \sum_{i=0}^{r-1}i t_i=\binom{r}{2}.
    \end{equation}
 Indeed, if 
$\sum_{i=0}^{r-1}i t_i\geq \binom{r}{2}+1$, then 
   \begin{align}
{l}\Big(\binom{r}{2}+1\Big)&> {l}\binom{r}{2}+r{d}\geq  \binom{r}{2}{l}+r{z}=\sum_{i=0}^{r-1}b_i \nonumber\\
&=\sum_{i=0}^{r-1}\sum_{b\in \cB\cap D_i}b\geq \sum_{i=0}^{r-1}\sum_{b\in \cB\cap D_i} i {l}=\sum_{i=0}^{r-1}it_i  {l} 
\geq {l}\Big(\binom{r}{2}+1\Big), \label{eq:>}
    \end{align}
    and we arrive at a contradiction. Similarly,  if $\sum_{i=0}^{r-1}i t_i\leq \binom{r}{2}-1$, then
  \begin{align}
    \binom{r}{2}{l}&< \binom{r}{2}{l}+r{z}=
    \sum_{i=0}^{r-1}\sum_{b\in \cB\cap D_i}b\leq \sum_{i=0}^{r-1}\sum_{b\in \cB\cap D_i} (i {l}+{d}) \nonumber\\
    & =\sum_{i=0}^{r-1}t_i( i {l}+{d})
    \leq 
    \Big(\binom{r}{2}-1\Big){l}+r{d}<\binom{r}{2}{l}, \label{eq:<}
   \end{align}
 and  \eqref{eq:<} makes no sense, and thus \eqref{eq:iti} holds.

Finally, recalling \eqref{eq:A}, let 
     $$
     \cB\cap D_i=\{i{l}+b_{i,j}, 1\leq j \leq t_i\}
     $$
where the $b_{i,j}$'s are $t_i$ distinct elements of $D$.  Then over $\mathbb{Z}$
    $$
    \binom{r}{2}{l}+r{z}=\sum_{i=0}^{r-1}b_i=\sum_{i=0}^{r-1}\sum_{j=1}^{t_i}(i{l}+b_{i,j})
=\binom{r}{2}{l}+\sum_{i=0}^{r-1}\sum_{j=1}^{t_i}b_{i,j},
       $$
hence 
   $$
   \sum_{i=0}^{r-1}\sum_{j=1}^{t_i}b_{i,j}=r{z}.
   $$ 
Now \eqref{eq:3ap} implies that $b_{i,j}={z}$ for all $i,j$. However, the numbers $b_i$ were chosen distinct, and thus, $t_i=1$ for all $i\le r.$ 
Moreover, $b_i=i{l}+{z}, i=0,1,\dots,r-1.$ On account of  \eqref{eq2} and \eqref{eq:Ai} the proof is complete.
 \qed

\vspace*{.15in}
\noindent{\em Proof of Lemma \ref{lemma:B}:} We closely follow \cite{Alon2001}, adding some details. Let $h$ be an integer, to be chosen later. Consider a set of integer numbers $D=(x_i)_i$ written 
in the form $x_i=\sum_{j=0}^t x_{i,j}h^j,$ where $0\le x_{i,j}<\frac hr, i=0,1,\dots,t,$ $t=\lfloor \log_hm\rfloor-1,$
and suppose further that for every $x_i\in D$
   $$
   \sum_{j=0}^t x_{i,j}^2=B.
   $$
If an $(r+1)$-tuple $x_0,x_1,\dots,x_{r+1}$ satisfies \eqref{eq:3ap}, then for every $j=0,1,\dots,t$
    \begin{equation}\label{eq:coor}
  x_{0,j}+x_{1,j}+\dots+x_{r-1,j}=rx_{r,j}.
  \end{equation}
By the convexity of the function $z\mapsto z^2$ this implies that
  $$
  x_{0,j}^2+x_{1,j}^2+\dots+x_{r-1,j}^2 \ge rx_{r,j}^2,
  $$
with equality if and only if $x_{0,j}=x_{1,j}=\dots=x_{r,j}.$ At the same time, 
  $$
  \sum_{i=0}^{r-1}\sum_{j=0}^t x_{i,j}^2=rB=r\sum_{j=0}^t x_{r,j}^2.
  $$
The last two relations imply that only identical $(r+1)$-tuples satisfy \eqref{eq:coor}, and thus only identical $(r+1)$-tuples of
elements in $D$ satisfy \eqref{eq:3ap}.
   
Clearly, $B\le (t+1) \frac {h^2}{r^2}$, so there is a choice of $B$ such that
    $$
    |D|\ge \frac {h^{t+1}}{r^{t+1}(t+1) \frac {h^2}{r^2}} \ge\frac{m}{h^3 r^{t-1}(t+1)}.
    $$
Take $h= \lfloor e^{\sqrt{\log m\log r}}\rfloor,$ then $(t-1)\log r<\sqrt{\log m\log r}$ and
   $$
   h^3 r^{t-1}(t+1)\le e^{5\sqrt{\log m\log r}}.\eqno{\qed}
   $$

\vspace*{.05in}
\subsection*{\sc Acknowledgments} Alexander Barg was partially supported by NSF grants CCF2110113 and CCF2104489. Itzhak Tamo was supported by the European Research Council (ERC grant number 852953) and by the Israel Science Foundation (ISF grant number 1030/15).


\begin{thebibliography}{1}

\bibitem{Alon2001}
N.~Alon.
\newblock Testing subgraphs in large graphs.
\newblock In {\em Proceedings 42nd IEEE Symposium on Foundations of Computer
  Science}, pages 434--441, 2001.

\bibitem{Behrend1946}
F.~A. Behrend.
\newblock On sets of integers which contain no three terms in arithmetical
  progression.
\newblock {\em Proc. Nat. Acad. Sci. U.S.A.}, 32:331--332, 1946.

\bibitem{cai2020construction}
H.~Cai, Y.~Miao, M.~Schwartz, and X.~Tang.
\newblock A construction of maximally recoverable codes with order-optimal
  field size.
\newblock arXiv:2011.13606, 2020.

\bibitem{Gabrys2019}
R.~Gabrys, E.~Yaakobi, M.~Blaum, and P.~H. Siegel.
\newblock Constructions of partial {MDS} codes over small fields.
\newblock {\em IEEE Transactions on Information Theory}, 65(6):3692--3701,
  2019.

\bibitem{Gopalan2014}
P.~Gopalan, C.~Huang, B.~Jenkins, and S.~Yekhanin.
\newblock Explicit maximally recoverable codes with locality.
\newblock {\em IEEE Trans. Inform. Theory}, 60(9):5245--5256, 2014.

\bibitem{Gopi2020}
S.~Gopi, V.~Guruswami, and S.~Yekhanin.
\newblock Maximally recoverable {LRC}s: a field size lower bound and
  constructions for few heavy parities.
\newblock {\em IEEE Trans. Inform. Theory}, 66(10):6066--6083, 2020.

\bibitem{Guruswami2020}
V.~Guruswami, L.~Jin, and C.~Xing.
\newblock Constructions of maximally recoverable local reconstruction codes via
  function fields.
\newblock {\em IEEE Transactions on Information Theory}, 66(10):6133--6143,
  2020.

\bibitem{Martinez2019}
U.~Mart{\'\i}nez-Pe{\~n}as and F.~R. Kschischang.
\newblock Universal and dynamic locally repairable codes with maximal
  recoverability via sum-rank codes.
\newblock {\em IEEE Trans. Inform. Theory}, 65(12):7790--7805, 2019.

\bibitem{Neri2020}
A.~Neri and A.-L. Horlemann-Trautmann.
\newblock Random construction of partial {MDS} codes.
\newblock {\em Des. Codes Cryptogr.}, 88(4):711--725, 2020.

\end{thebibliography}
\end{document}